\documentclass[showpacs,preprintnumbers,amsmath,amssymb]{revtex4}
\usepackage{epsfig}
\usepackage{psfrag}
\usepackage{amsfonts}
\usepackage{graphicx}% Include figure files
\usepackage{dcolumn}% Align table columns on decimal point
\usepackage{bm}% bold math
\usepackage{xcolor}
\usepackage{amsmath}
\usepackage{amssymb}
\usepackage{amsthm}
\usepackage[mathscr]{eucal}

\begin{document}

\title{Absence of Landau-Peierls Instability in the Magnetic Dual Chiral Density Wave Phase of Dense QCD}

\author{E. J. Ferrer and V. de la Incera}

\affiliation{Department of Physics and Astronomy, University of Texas Rio Grande Valley, 1201 West University Dr., Edinburg, TX 78539}

\date{\today}

\begin{abstract}
We investigate the stability of the Magnetic Dual Chiral Density Wave (MDCDW) phase of cold and dense QCD against collective low-energy fluctuations of the order parameter. The appearance of additional structures in the system free-energy due to the explicit breaking of the rotational and isospin symmetries by the external magnetic field play a crucial role in the analysis. The new structures stiffen the spectrum of the thermal fluctuations in the transverse direction, thereby avoiding the Landau-Peierls instability that affect single-modulated phases at arbitrarily low temperatures. The lack of Landau-Peierls instabilities in the MDCDW phase makes this inhomogeneous phase of dense quark matter of particular interest for the physics of neutron stars.
\end{abstract}

                             
\maketitle

\section{Introduction}
In recent years, a great deal of efforts has been dedicated to complete the temperature-density phase map of QCD. The regions of extreme high temperature or density are better understood in part thanks to the weakening of the strong coupling by the phenomenon of asymptotic freedom.  They are described by the quark-gluon plasma (QGP) phase at high temperature and low density, or by the color superconducting color-flavor locked (CFL) phase at asymptotically large density and low temperature \cite{QCDreviews}. More challenging, nonetheless, is to determine the phases that take place at intermediate density-temperature regions, where lattice QCD is not applicable due to the sign problem, so one has to rely on nonperturbative methods and effective theories. 

It has long been argued that the region of intermediate density and relatively low temperature may feature inhomogeneous phases, many of which have spatially inhomogeneous chiral condensates that are favored over the homogeneous ones. Such spatially inhomogeneous phases have been found in the large-N limit of QCD \cite{largeNQCD,largeN}, in NJL models \cite{DCDW}-\cite{PRD85-074002}, and in quarkyonic matter \cite{q-chiralspirals}. In all the cases, single-modulated chiral condensates are energetically favored over chiral condensates with higher-dimensional modulations.

However, single-modulated phases in three spatial dimensions are known to be unstable against thermal fluctuations at any arbitrarily small finite temperature, a phenomenon known in the literature as Landau-Peierls instability \cite{Landau-Peirls Inst}. In dense QCD models, the Landau-Peierls instability occurs in the periodic real kink crystal phase \cite{Hidaka}; in the Dual Chiral Density Wave (DCDW) phase \cite{Tatsumi}; and in the quarkyonic phase \cite{Pisarski}. The instability signals the lack of long-range correlations at any finite temperature and hence the lack of a true order parameter. Only a quasi long-range order remains in all these cases, a situation that resembles what happens in smectic liquid crystals \cite{smectic liquid crystal}.

It is worth to notice, on the other hand, that the two scenarios where dense quark matter phases can be realized, neutron stars \cite{Nucleons-B, Quark-B} and heavy-ion collisions \cite{B-HIC}, typically have very strong magnetic fields. Consequently, the field effect on dense quark matter phases has become a hot topic of investigation (see \cite{Lec-Notes,EPJ_A52} and references therein), adding an extra dimension to the QCD phase map problem. The effect of a magnetic field in inhomogeneous phases has been considered in quarkyonic matter \cite{ferrer-incera-sanchez}, and in the DCDW phase\cite{KlimenkoPRD82, PLB743}. 

The presence of a magnetic field is relevant due to the activation of new channels of interaction and, occasionally, the generation of additional condensates. For instance, in the quarkyonic phase, a magnetic field is responsible for the appearance of a new chiral spiral between the pion and magnetic moment condensates, $\langle\bar{\psi}\gamma^5\psi\rangle$ and $\langle \bar{\psi}\gamma^1\gamma^2\psi\rangle$ respectively, \cite{ferrer-incera-sanchez}.  Additional condensates also emerge in the homogeneous chiral phase \cite{AMM}, as well as in color superconductivity \cite{ferrer-incera}.

Adding a magnetic field to the DCDW phase explicitly reduces the rotational and isospin symmetries, significantly enhances the window for inhomogeneity \cite{KlimenkoPRD82, PLB743}, and leads to topologically nontrivial transport properties \cite{Topological-Transport-1, Topological-Transport-2} related to the spectral asymmetry of the lowest Landau level (LLL). Because the DCDW phase in a magnetic field is physically distinguishable from the conventional DCDW one at zero field, it has been termed the Magnetic DCDW (MDCDW) phase \cite{Topological-Transport-1}.

In this paper, we go beyond the mean-field approximation to investigate the stability of the MDCDW phase against thermal fluctuations in the region of relevance for neutron star applications. As it will be shown below, an external magnetic field introduces new structures in the system's free energy that make it anisotropic, so that  terms with transverse and parallel (to the magnetic field direction) derivatives of the order parameter enter with different coefficients. The field-induced structures also affect the condensate minimum equations. Furthermore, the coefficients of these structures contribute to the free energy of the fluctuations in such a way that their dispersion becomes linear in all the directions and thus the system lacks the severe infrared divergencies that characterize the Landau-Peierls instability. This means that the long-range order is not wiped out by thermal fluctuations at arbitrarily low temperatures, in nice agreement with the general arguments presented in \cite{Topological-Transport-2} for the MDCDW phase and in \cite{Hidaka} for single-modulated phases with an external vector field.

The paper is organized as follow. In Section II, the generalized Ginzburg-Landau (GL) expansion of the MDCDW thermodynamic potential in powers of the order parameter and its derivatives is obtained and shown to differ from the corresponding DCDW expansion due to two types of new structural terms induced by the magnetic field. The new structural terms make the GL expansion anisotropic and modify the stationary equation that determine the ground state of the system. In Section III, we derive the low-energy theory of the thermal fluctuations about the condensate solution, demonstrating how the magnetic field lead to a spectrum of thermal fluctuations that is linear in all the directions.  We then show how this property prevents the system to exhibit the Landau-Peierls instability despite being a single-modulated phase. The origin and physical interpretations of the new terms, as well as the implications of our findings are discussed throughout the paper and summarized in our concluding remarks in Section IV.

  \section{Low-Energy Ginzburg-Landau Expansion in the MDCDW Phase}
 
The MDCDW phase emerges in a two-flavor NJL theory of interacting quarks at finite baryon density in the presence of a magnetic field \cite{KlimenkoPRD82}. The ground state of this phase is characterized by an inhomogeneous chiral condensate 
\begin{equation}\label{condensate}
\langle\bar{\psi}\psi\rangle+i\langle\bar{\psi}i\gamma^5\tau_3\psi\rangle=\bar{\Delta}\exp(i\bar{q}z)=-\frac{1}{2G}M_0(z), 
\end{equation}
identical in form to the one that exists in the single-modulated DCDW phase, except that the modulation vector there can be arbitrarily chosen along any direction, while in the MDCDW a modulation vector parallel to the magnetic field is energetically favored. Without loss of generality, we choose the magnetic field $\mathbf{B}$ in the z-direction and thus the condensate modulation is $\mathbf{q}=(0,0,\bar{q})$. 

The two-flavor NJL theory in the absence of a magnetic field exhibits the following global symmetries: vector and axial isospin $SU_V(2)\times SU_A(2)$, full spatial rotations $SO(3)$, and translations $R^3$. However, when an external $\mathbf{B}$ is present, it explicitly breaks the isospin symmetry and the rotations about the axes perpendicular to the field direction. Hence, in this case the symmetry reduces to $U_V(1)\times U_A(1)\times SO(2)\times R^3$. 

The low-energy theory of the MDCDW phase is described by a generalized Ginzburg-Landau (GL) expansion of the thermodynamic potential in powers of the order parameter and its derivatives. The GL expansion usually assumes that both the order parameter and its derivatives are small, a condition that may occur near the phase transition. The GL expansion of the MDCDW phase near the critical point (CP), that is, in the region of large temperatures and low chemical potentials, was explored in \cite{PLB743}. For neutron star applications, however, the region of interest is that of large chemical potentials and low temperatures. Henceforth, our study will be mainly focused on the region of relevance for neutron stars near the transition to the chirally restored phase. Later in this section, we shall discuss the validity of the GL expansion in this region.

The GL expansion must be invariant under the symmetries of the theory in the presence of the external magnetic field. In the MDCDW system, the order parameter is characterized by the scalar and pseudoscalar fields $\sigma= -2G\bar{\psi}\psi$ and $\pi=-2G\bar{\psi} i\gamma^5\tau_3\psi$, respectively. Under a global chiral transformation $e^{i\gamma_5 \tau_3\theta/2}$ of the fermion fields, they transform as $\sigma \to \sigma \cos \theta+\pi \sin \theta$ and $\pi \to \pi \cos \theta- \sigma \sin \theta$, reflecting the isomorphism between the chiral group  $U_A(1)$ and the $SO(2)$ of internal rotations acting on the two-dimensional vector $\phi=(\sigma, \pi)$. In a similar way, one can see that the $U_V(1)$ transformations of the fermions reduce to the trivial group acting on the vector $\phi$. 

Using the SO(2) representation, the GL expansion can be written as
\begin{eqnarray}\label{GL-Free Energy-phi}
\mathcal{F}&=&a_{2,0}\phi^T\phi+\frac{b_{3,1}}{2}\left [\phi^T\hat{B}\cdot \widetilde{\nabla} \phi+\hat{B}\cdot (\widetilde{\nabla} \phi)^T \phi\right ]+a_{4,0} (\phi^T\phi)^2\nonumber
\\
&+&a^{(0)}_{4,2}(\widetilde{\nabla}\phi)^T\cdot \widetilde{\nabla}\phi+a^{(1)}_{4,2} \hat{B}\cdot (\widetilde{\nabla} \phi)^T \hat{B}\cdot \widetilde{\nabla} \phi \nonumber
\\
&+&\frac{b_{5,1}}{2}(\phi^T\phi)\left [\phi^T\hat{B}\cdot \widetilde{\nabla} \phi+\hat{B}\cdot (\widetilde{\nabla} \phi)^T \phi \right ]+ \frac{b_{5,3}}{2}\left [( \widetilde{\nabla}^2 \phi)^T\hat{B}\cdot \widetilde{\nabla}\phi+\hat{B}\cdot (\widetilde{\nabla}\phi)^T \widetilde{\nabla}^2 \phi \right ]+a_{6,0}(\phi^T\phi)^3 \nonumber
\\
&+&a^{(0)}_{6,2}(\phi^T\phi)(\widetilde{\nabla}\phi)^T\cdot \widetilde{\nabla}\phi+a^{(1)}_{6,2}(\phi^T\phi) [\hat{B}\cdot (\widetilde{\nabla} \phi)^T \hat{B}\cdot \widetilde{\nabla} \phi]+a_{6,4}(\widetilde{\nabla}^2 \phi)^T(\widetilde{\nabla}^2 \phi)+... ,
\end{eqnarray}
where we introduced the normalized vector $\hat{B}=\mathbf{B}/|\mathbf{B}|$ in the direction of the magnetic field and used it to form additional structural terms that are consistent with the symmetry of the theory in a magnetic field. In this representation the gradient operator is given by
\begin{equation}
\widetilde{\nabla}= \left(\begin{array}{ccc}0 & 1 \\-1 & 0\end{array}\right)\nabla \qquad .
\end{equation}

The coefficients $a$ and $b$ are functions of $T, \mu$ and $B$. They can be derived from the MDCDW thermodynamic potential found in \cite{Topological-Transport-2}, although their explicit expressions are not relevant for the present study. The first subindex in the coefficient's notation indicates the power of the order parameter plus its derivatives in that term, the second index denotes the power of the derivatives alone.

There are two distinguishable effects of the external magnetic field in the GL expansion. On the one hand, the terms with coefficients $a^{(1)}_{i,j}$ are responsible for the explicit separation of transverse and parallel derivatives, as it is expected to occur in any theory where the rotational symmetry is broken by an external vector. These terms are similar to those with coefficients $a^{(0)}_{i,j}$,  except that the gradient is replaced by the projection of the gradient along the external field. These additional structures were missed in previous studies of the MDCDW low-energy theory \cite{PLB743,Yoshiite}, despite they affect the fluctuation spectrum, a fact that will become apparent in the next section. On the other hand, the terms with  coefficients $b_{i,j}$ are not a common feature of theories with an external vector but rather a consequence of the existence of a nontrivial topology in the system. In the MDCDW case, the nontrivial topology is due to the spectral asymmetry of the Lowest Landau Level (LLL) fermions \cite{KlimenkoPRD82}.

The isomorphism between SO(2) and $U_A(1)$ allows to represent the order parameter as a complex function $M(x)=\sigma(x)+ i \pi(x)$. In terms of $M(x)$ the GL expansion of the free energy (\ref{GL-Free Energy-phi}) takes the form
\begin{eqnarray}\label{GL-Free Energy}
\mathcal{F}&=&a_{2,0}|M|^2-i\frac{b_{3,1}}{2}\left [M^*(\hat{B}\cdot \nabla M)-(\hat{B}\cdot \nabla M^*)M\right ]+a_{4,0} |M|^4+a^{(0)}_{4,2}|\nabla M|^2
\nonumber
\\
&+&a^{(1)}_{4,2} (\hat{B}\cdot \nabla M^*)(\hat{B}\cdot \nabla M)-i\frac{b_{5,1}}{2}|M|^2\left [M^*(\hat{B}\cdot \nabla M)-(\hat{B}\cdot \nabla M^*)M\right ]\nonumber
\\
&+&\frac{ib_{5,3}}{2}\left[(\nabla^2M^*) \hat{B}\cdot \nabla M- \hat{B}\cdot \nabla M^*(\nabla^2 M)\right]+a_{6,0}|M|^6+a^{(0)}_{6,2}|M|^2|\nabla M|^2\nonumber
\\
&+&a^{(1)}_{6,2}|M|^2(\hat{B}\cdot \nabla M^*)(\hat{B}\cdot \nabla M)+a_{6,4}|\nabla^2 M|^2 +... 
\end{eqnarray}

Assuming that the MDCDW order parameter is a single-modulated density wave $M(z)=m e^{iqz}$, with $m \equiv -2G \Delta$, the free energy (\ref{GL-Free Energy}) becomes
\begin{eqnarray}\label{GL-Free Energy-qdelta}
\mathcal{F}&=&a_{2,0}m^2+b_{3,1}qm^2+a_{4,0} m^4+a_{4,2}q^2m^2+b_{5,1}qm^4\nonumber
\\
&+&b_{5,3}q^3m^2+a_{6,0}m^6+a_{6,2}q^2m^4+a_{6,4}q^4m^2,
\end{eqnarray}
where $a_{4,2}=a^{(0)}_{4,2}+a^{(1)}_{4,2}$, $a_{6,2}=a^{(0)}_{6,2}+a^{(1)}_{6,2}$, and we kept up to sixth order terms to ensure the stability of the MDCDW phase in the mean-field approximation. 

We can now explain why the expansion (\ref{GL-Free Energy-qdelta}) is valid in the region $T < m_{V}<\mu$, near the chirally restored phase, with $m_V$ the vacuum quark mass. In such a region, the order parameters satisfy $m/\mu<1$  and $q/2\mu<1$ \cite{KlimenkoPRD82}. One can readily show \cite{footnote}, following an approach similar to the one used in \cite{PRD97-036009} for the DCDW case, that the power series in $q$ effectively becomes an expansion in powers of $q/2\mu$, hence corroborating the consistency of the expansion and the truncation used. 

Straight derivations \cite{footnote} show that the $a$ coefficients get contributions from all Landau levels $l$. In contrast, the higher Landau levels (hLL) $l>1$, do not contribute to the $b$-coefficients. This can be understood from the following observations: the $b$-terms in (\ref{GL-Free Energy-qdelta}) are odd in q, but, as can be gathered from the fermion spectrum (Eqs. (9) and (10) in \cite{Topological-Transport-2}) and the thermodynamic potential (Eq. (70) in \cite{Topological-Transport-2}), the part of the potential that comes from the hLL is invariant under the change $q\to -q$, so it cannot generate odd-in-q terms in the GL expansion. That leaves the LLL modes as the only possible source of the $b$-terms. Indeed, the LLL contribution is not invariant under $q\to-q$, due to the asymmetry of the LLL modes. In principle, the LLL part of the thermodynamic potential can have q-odd and q-even terms. Obviously the $b$-terms come from the odd part. Such an odd-part is topological in nature, a fact that manifests in the existence of several anomalous quantities, like the anomalous part of the quark number, which is proportional to a topological invariant \cite{PLB743}, or the anomalous electric charge and the anomalous Hall current  \cite{Topological-Transport-2}, all of which are odd in $q$. 

So in summary, the additional $a$ and $b$ terms have quite different origins. New $a$ terms will always appear if an external vector is added to the system because they simply reflect the explicit breaking of the rotational symmetry produced by this vector. The $b$ terms, however, come from the topology of the modified fermion spectrum in the ground state with the external field.  As the LLL part of the thermodynamic potential is linear in the magnetic field $B$, so will be the $b$-coefficients.
 
The ground state is found as the solution of the stationary equations
\begin{subequations}
\begin{align}
\partial \mathcal{F}/\partial m&=&2m \{ a_{2,0}+2a_{4,0} m^2+3a_{6,0}m^4+q^2[a_{4,2}+2a_{6,2}m^2+a_{6,4}q^2]
+q [b_{3,1}+2b_{5,1} m^2+b_{5,3}q^2]\}=0
\label{SC-Delta}
\\
\partial \mathcal{F}/\partial q&=&m^2  \{2q[a_{4.2}+a_{6,2}m^2+2a_{6.4}q^2] +b_{3,1}+b_{5,1} m^2+3b_{5,3}q^2\}=0 \qquad \qquad \qquad\qquad \qquad\quad \qquad
\label{SC-q}
\end{align}
\end{subequations}

It is easy to see that in the limit of zero-magnetic-field, where the $a^{1}_{i,j}$ and $b_{i,j}$ coefficients vanish, one recuperates the minimum equations of the DCDW phase \cite{Tatsumi}, as it should have been expected.

\section{No Landau-Peierls Instability at $B \neq 0$}

To explore the Landau-Peierls instability we need to go beyond the mean-field approximation and investigate the low-energy thermal fluctuations that may affect the long-range order of the inhomogeneous ground state. In principle, there can be fluctuations of the condensate magnitude and of the condensate phase. However, not all of them are associated to the spontaneous breaking of a global symmetry. We notice that to probe the instability of the ground state at arbitrarily low temperatures, the relevant fluctuations are those that can be excited at very low energies, i.e., those generated by the Goldstone bosons of the system. Hence, in our analysis, we do not consider the magnitude fluctuations because they are not associated to a Goldstone mode.

The ground state of the MDCDW system spontaneously breaks the chiral symmetry $U_A(1)$ and the translation along z, reducing the symmetry group to  $U_V(1)\times SO(2) \times R^2$. Hence, there are, at least in principle, two Goldstone bosons: the neutral pion $\tau$ and the phonon $\xi$. Before considering their fluctuations, it is convenient to examine the effect of the global transformations of these broken groups on the order parameter, 
\begin{equation}\label{SB-Pattern-Finite}
M(x) \rightarrow e^{i\tau}M(z+\xi)=e^{i(\tau+q\xi)}M(z)
 \end{equation}

From Eq. (\ref{SB-Pattern-Finite}), one clearly sees that there is a locking between the chiral rotation and the z-translation. Therefore, we can always express them as two orthogonal combinations, one that leaves the order parameter invariant and one that changes it. As a consequence, there is only one legitimate Goldstone field in the MDCDW theory. One can arbitrarily choose it as either the pion, the phonon, or a linear combination of them. Henceforth, without loss of generality, we consider it to be the phonon. 

We now subject the order parameter to a small phonon fluctuation $u(x)$ and expand it about the condensate solution up to quadratic order in the fluctuation, 
\begin{equation}\label{Fluctuation}
M(x)=M(z+u(x))\simeq M_0(z)+M'_0(z)u(x)+\frac{1}{2} M''_0(z) u^2(x),
 \end{equation}
where $M_0(z)=\bar{m}e^{i\bar{q}z}$ is the ground state (\ref{condensate}) with $\bar{m}$ and $\bar{q}$ the solutions of (\ref{SC-Delta})-(\ref{SC-q}). 

Inserting (\ref{Fluctuation}) into (\ref{GL-Free Energy}), and keeping terms up to quadratic order in $u(x)$, we arrive at the phonon free energy 
\begin{equation}\label{Fluctuation-Free Energy}
\mathcal{F}[M(x)]=\mathcal{F}_0+v^2_z(\partial_z\theta)^2+v_\bot^2(\partial_\bot \theta)^2+\zeta^2(\partial_z^2\theta+\partial_\bot^2\theta)^2,
\end{equation} 
where we rewrote the free energy in terms of a dimension-1 pseudo boson field $\theta=qmu$, and introduced the notation
$\mathcal{F}_0=\mathcal{F}(M_0)$, $(\partial_\bot \theta)^2=(\partial_x \theta)^2+(\partial_y\theta)^2$ and $\zeta^2=a_{6.4}$.

The coefficients $v^2_z$, $v_\bot^2$ are the squares of the parallel and transverse group velocities respectively given by

\begin{equation}\label{Parallel-Coef2}
v^2_z= a_{4.2}+m^2 a_{6.2}+6q^2a_{6.4}+3qb_{5,3}
\end{equation}
\begin{equation}\label{Transverse-Coef2}
v_\bot^2= a_{4.2}+m^2 a_{6.2}+2q^2 a_{6.4}+qb_{5,3}-a^{(1)}_{4.2}-m^2a^{(1)}_{6.2}
\end{equation}  

For the sake of simplicity, we dropped the bar notation for the dynamical parameters $m$ and $q$ in (\ref{Parallel-Coef2}) and (\ref{Transverse-Coef2}), but they must be understood as the solutions of the stationary conditions (\ref{SC-Delta})-(\ref{SC-q}). In deriving (\ref{Fluctuation-Free Energy}), the term linear in $\partial_z \theta$ cancels out after considering (\ref{SC-q}).

The corresponding low-energy Lagrangian density is
\begin{equation}\label{phononLagrangian}
\mathcal{L}=\frac{1}{2}[(\partial_0\theta)^2-v^2_z(\partial_z\theta)^2-v_\bot^2(\partial_\bot \theta)^2-\zeta^2(\partial_z^2\theta+\partial_\bot^2\theta)^2],
\end{equation}

The resultant phonon spectrum is anisotropic and linear in the longitudinal and transverse directions
\begin{equation}\label{spectrum}
E\simeq\sqrt{v^2_zk^2_z +v_\bot^2k_\bot^2},
\end{equation}
where $k_\bot^2=k_x^2+k_y^2$.

The spectrum of the gapless excitations over the same background was also worked out in \cite{Yoshiite}. However, a direct comparison of the results is not possible because there the structures with coefficients $a^{(1)}_{i,j}$ were not considered in the GL expansion and besides, their free-energy included the magnitude fluctuations which are irrelevant to probe the Landau-Peierls instability. We underline that equations (\ref{Transverse-Coef2}) and (\ref{spectrum}) highlight why the $a^{(1)}_{i,j}$ coefficients cannot be ignored in the analysis, as they explicitly affect the phonon spectrum and through it, the behavior of the thermal fluctuations.

Notice that in the absence of the magnetic field, the coefficients $a^{(1)}$ and $b$ vanish and, at the same time, the remaining combination in (\ref{Transverse-Coef2}) becomes zero due to the stationary condition (\ref{SC-q}), leading to $v_\bot=0$. On the other hand, $v_z\neq 0$ because $a_{6.4}$ cannot be zero for the minimum solution to exist \cite{NickelPRD80}. As a consequence, the spectrum becomes soft in the transverse direction. This is precisely the origin of the Landau-Peierls instability in the DCDW phase, as will become clear in the discussions below. 

To investigate the stability of the condensate against the fluctuations we calculate its average
\begin{equation}\label{averageM}
\langle M \rangle=m e^{iqz}\langle \cos qu \rangle,
\end{equation}
where the average is defined as 
 \begin{equation}\label{average}
\langle  ... \rangle =\frac{\int \mathcal{D}u(x) ... e^{-S(u^2)}}{\int \mathcal{D}u(x) e^{-S(u^2)}}
\end{equation}
with
 \begin{equation}\label{phonon action}
S(u^2)= T\sum_n \int^{\infty}_{-\infty} \frac{d^3k}{(2\pi)^3} [\omega^2_n+(v^2_zk^2_z +v_\bot^2k_\bot^2+\zeta^2 k^4)]q^2m^2u^2.
\end{equation}
the finite-temperature effective action of the phonon and $\omega_n=2n\pi T$ is the Matsubara frequency. 

We can now consider the relation
\begin{equation}\label{cos-exp-relation}
\langle \cos qu \rangle =e^{-\langle (qu)^2 \rangle/2}
\end{equation}
and use (\ref{average}) to find the mean square of the fluctuation as
\begin{eqnarray}\label{Fluctuation-2}
\langle q^2u^2 \rangle &=&\frac{1}{(2\pi)^2}\int_{0}^\infty dk_\bot k_\bot \int_{-\infty}^\infty dk_z \frac{T}{m^2(v^2_zk^2_z +v_\bot^2k_\bot^2+\zeta^2 k^4)}\nonumber
\\
&\simeq& \frac{\pi T}{m\sqrt{v^2_zv_\bot^2}}.
\end{eqnarray}
where we took into account that the lowest Matsubara mode is dominant in the infrared. 

The first thing to notice from (\ref{Fluctuation-2}) is that $\langle q^2u^2 \rangle$ is finite then $\langle M \rangle \neq 0$. This means that at $B \neq 0$ the fluctuations do not automatically wipe out the condensate at arbitrarily low $T$, or in other words, the MDCDW system does not exhibit the Landau-Peierls instability. 

This finding is in sharp contrast with the case at zero magnetic field, where 
 \begin{equation}\label{Fluctuation-3}
\langle u^2 \rangle =\frac{1}{(2\pi)^2}\int_{l_\bot^{-1}}^\infty dk_\bot k_\bot \int_{-\infty}^\infty dk_z \frac{T}{v^2_zk_z^2+\zeta^2k_\bot^4} \simeq \frac{T}{4\pi v_z\zeta}   \ln \left( \frac{v_zl_\bot}{\zeta} \right),
 \end{equation}
is infrared divergent when $l_\bot \to \infty$, leading to $\langle M \rangle= 0$ at any nonzero $T$. 

Comparing the two results, one realizes that the lack of Landau-Peierls instabilities in the presence of a magnetic field is due to the stiffening of the spectrum in the transverse direction. As discussed above, this feature is a direct consequence of the explicit breaking of the rotational symmetry by the external field, which is the necessary condition for the coefficients $a^{(1)}_{i,j}$ to be turned on. Therefore, this behavior should happen in spatially inhomogenous systems, even if they have no nontrivial topology in their fermion structure, i.e., in cases where the $b_{i,j}$-type coefficients are not present.

A second important point to highlight from the result (\ref{Fluctuation-2}) is that as the temperature increases and the system approaches the transition to the chirally restored phase, $m \to 0$ and the fluctuations tend to exponentially wipe out the long-range order. The threshold temperature $T_h$ at which this regime becomes operative can be found from the condition $T_h=m |v_z| |v_\bot|/\pi$. If $T_h < T_c $, with $T_c$ the temperature for chiral restoration, there is a region of temperatures $T_h< T< T_c$ where despite the presence of the magnetic field, the long-range order is effectively wiped out by the phonon thermal fluctuations. In this region the system will likely be characterized by a quasi-long-range order \cite{Hidaka} similar to the smectic phase of liquid crystals \cite{smectic liquid crystal}.

It is worthy to notice that the lack of Landau-Peierls instabilities in the presence of a magnetic field will not be changed by a nonzero current quark mass, since this property comes from the effect of the magnetic field on the low energy behavior of the phonon, which persists as a Goldstone boson even at nonzero quark masses.

\section{Conclusions}
In this paper we have shown that a magnetic field eliminates the Landau-Peierls instability of the single modulated DCDW phase in dense QCD. In this context, the magnetic field plays a dual role. On the one hand, it acts as an extra vector that explicitly breaks the rotational and isospin symmetries, allowing the formation of additional structures in the GL expansion of the MDCDW thermodynamic potential and making it anisotropic. On the other hand, it induces a nontrivial topology in the system that manifests itself in the asymmetry of the LLL modes and in the appearance of odd-in-q terms in the GL expansion. These two features in turn affect the low-energy theory of the thermal fluctuations, giving rise to a linear transverse mode in the fluctuation dispersion relation, thereby preventing the Landau-Peierls instability, which is a hallmark of single-modulated phases in $3+1$ dimensions at finite temperature. 

Physically, the new terms correspond to new channels of interactions expressed as powers of the coupling of the magnetic field with the magnetic moment of the order parameter $\sim\hat{B}\cdot \nabla M$. The order parameter has a magnetic moment because it is a neutral composite scalar, made of oppositely charged fermions with opposite spins. From the new interaction channels, those with $a^{(1)}_{i,j}$ coefficients have even powers of $\sim\hat{B}\cdot \nabla M$ and hence get contributions from both, the hLL and the even-in-$q$ part of the LLL. Meanwhile, odd-power terms have a topological origin that can be traced back to the asymmetry of the LLL and its odd-in-$q$ contributions to the thermodynamic potential. This same feature is at the heart of the anomalous properties of the MDCDW phase. A  similar type of topological contribution was found in \cite{SON-Stephanov} for a neutral Goldstone boson $\pi_0$ that coupled to the magnetic field via the triangle anomaly.

The results obtained in this paper were derived taking into account that the internal symmetry group of the theory in a magnetic field is reduced from $SU(2) _V \times SU(2)_A$ to  $U(1) _V \times U(1)_A$, which allows to ignore the charged pions in the analysis. In a magnetic field, the charged pions acquire a field-induced mass $m^2_{\pi} \sim 2\widetilde{e}|\mathbf{B}|$ that increases with the field, even in the chiral limit. That is why the charged pions are not relevant for the existence of the Landau-Peierls instabilities since for that only gapless fluctuations are important. 

Nevertheless, one could expect that the charged pions could affect the threshold temperature $T_h$. This can be understood by noticing that at weak magnetic fields the mass of the charged pions is small, so they would behave as quasi Goldstone bosons and their fluctuations could in principle become relevant with increasing temperatures. However, ignoring the charged pions is self-consistent when their mass is larger than twice the mass of the quasiparticles, $m_{\pi} \gtrsim 2m$, as in this case they become unstable and decay into quasiparticle-quasihole pairs \cite{Phases}. Therefore,  $T_h$ can be found ignoring the charged pions, as long as the field is large enough to ensure this condition is met. At $T=0$, in the region of intermediate to high density where the condensate is favored over the chirally restored phase, the magnitude of the condensate sharply decreases with the chemical potential at strong \cite{KlimenkoPRD82} and weak \cite{Israel} coupling. Inserting a small temperature into the mix does not significantly change this trend. We can then conservatively (over)-estimate the field magnitude from the results at zero temperature. For example, quasiparticle masses of order $0.2 m_V\simeq$ 60MeV and smaller are reasonable estimates \cite{KlimenkoPRD82} for this region of chemical potential, so fields of order $10^{16}$G and higher should be enough to disregard the charged pions effects. Such a strength of magnetic field is reasonable for neutron star cores \cite{Quark-B,1903.08224}.

Although more work is still needed to determine the range of temperatures where the MDCDW phase remains stable for reasonable values of density and magnetic field, the lack of Landau-Peierls instability and the fact that this phase has been shown to be compatible with the observed 2$M_{\bigodot}$ mass of neutron stars \cite{InhStars}, make the MDCDW phase a robust candidate for the core of compact objects.

Acknowledgments: We are grateful for discussions with Sarang Gopalakrishnan, William Gyory and Vadim Oganesyan.  This work was supported in part by NSF grant PHY-2005331.


\begin{thebibliography}{00}
\bibitem{footnote} Details of the expansion coefficients will appear on a separate publication currently under preparation.
\end{thebibliography}
\end{document}